\documentclass[10pt,showpacs,twocolumn,nofootinbib,aps,prd]{revtex4-1}
\usepackage{CJK}
\usepackage{amsmath,amssymb,bm,mathrsfs}
\usepackage{mathtools,mathptmx,array,slashed}
\usepackage{graphicx,graphics,subfigure}
\usepackage{ctable,multirow}
\usepackage{tabularx}
\usepackage[colorlinks=true,linkcolor=blue,citecolor=blue,urlcolor=blue]{hyperref}

\DeclareSymbolFont{symbols} {OMS}{cmsy}{m}{n}

\def\be{\begin{equation}}
\def\ee{\end{equation}}
\def\bea{\begin{eqnarray}}
\def\eea{\end{eqnarray}}
\def\ba{\begin{aligned}}
\def\ea{\end{aligned}}

\def\pp{\partial}

\newcommand{\meq}[1]{(\ref{#1})}

\begin{document}

%\begin{CJK*}{GBK}{song}

\title{Light rings and shadows of static black holes in effective quantum gravity}

\author{Wentao Liu$^{1}$}

\author{Di Wu$^{2}$}
\email{wdcwnu@163.com}
% https://orcid.org/0000-0002-2509-6729

\author{Jieci Wang$^{1}$}
\email{jcwang@hunnu.edu.cn}

\affiliation{$^{1}$Department of Physics, Key Laboratory of Low Dimensional Quantum Structures and Quantum Control of Ministry of Education, and Synergetic Innovation Center for Quantum Effects and Applications, Hunan Normal University, Changsha, Hunan 410081, People's Republic of China \\
$^{2}$School of Physics and Astronomy, China West Normal University, Nanchong, Sichuan 637002, People's Republic of China}

\begin{abstract}
Recently, two types of static black hole models that retain general covariance have been proposed within the Hamiltonian constraint approach to effective quantum gravity (EQG).
We have studied the light rings and shadows of these black holes using the topological method and the backward ray-tracing method, respectively.
We demonstrate that these light rings in both types of static black holes are standard and unstable according to the classification of light rings.
Subsequently, we checked the position of the light rings using the photon trajectory equation.
We found that although the quantum parameters do not affect the light rings of these two types of black holes, they do reduce the size of the first type of static black hole in EQG, making it smaller.
However, for the second type of static black hole in EQG, we cannot distinguish it from a Schwarzschild black hole based on the shadow alone.
Fortunately, the quantum parameters shrink the lensing rings of both types of black holes in EQG, causing the black hole shadow to occupy a larger proportion within the ring.
This can serve as a basis for distinguishing whether the black hole is in EQG or general relativity (GR).
\end{abstract}
\maketitle

%\end{CJK*}

%%%%%%%%%%%%%%%%%%%%%%%%%%%
\section{Introduction}
%%%%%%%%%%%%%%%%%%%%%%%%%%%
Since the Event Horizon Telescope (EHT) Collaboration produced black hole images of M87* \cite{APJL875-L1,APJL875-L6} and Sgr A* \cite{APJL930-L12,APJL930-L14}, the light ring and black hole shadow have been considered promising tools to estimate the parameters of black holes, such as mass, spin, and the external fields in the environment \cite{PRD79-083004,
PRD100-024018,PRD102-024004,JHEP0720054,PRD103-044057,PRD104-044028,PRD106-064058,CQG40-165007}.
Recently, topology, as an effective mathematical tool to explore the properties of black holes, has received considerable interest and enthusiasm.
To date, there are two main facets to topology research.
One facet of the investigation focuses on the thermodynamic properties of black holes, such as thermodynamic phase transitions \cite{PRD105-104003,PRD107-044026,JHEP0623115,JHEP1123068,PLB854-138722,2404.02526,
JHEP0324138,JHEP0624213} and thermodynamic topological classification \cite{JHEP0624213,PRL129-191101,PRD107-064023,JHEP0123102,
PRD107-024024,PRD107-084002,EPJC83-365,EPJC83-589,PRD108-084041,PLB856-138919,2404.08243,
2409.09333,2409.11666,2409.12747}.
Another facet of investigation focuses on the light rings \cite{PRL119-251102,
PRL124-181101,PRD102-064039,PRD103-104031,PRD104-044019,PRD105-024049,PRD105-064070,
PRD108-104041,PRD109-064050,AP156-102920,AP162-102994,2405.18798} of some black holes and has been extended to timelike circular orbits \cite{PRD107-064006,JCAP0723049,2406.13270}, which may provide more footprints for the observation of black holes.
Furthermore, the study of black hole shadows can provide insights into several fundamental problems, including the extreme environments around black holes, dark matter, the accelerating universe, extra dimensions and quantum effects of gravity.
Examples of such investigations can be found in various studies, as highlighted in Refs. \cite{PRD89-124004,PRD98-084063,PRD100-044055,PRD100-044057,PRD100-024020,
JHEP1019269,EPJC80-790,PRD104-064039,PRD103-064026,PRD103-104033,EPJC81-991,APJ916-116,
APJ938-2,CTP74-097401,SCPMA65-290411,EPJC82-835,SCPMA66-110411,JCAP1122006,APJ957-103,
APJ958-114,EPJC83-619,2404.12223,JCAP0124059,APJ954-78,CQG40-174002,PRD109-064027,
2307.16748,PRD109-124062,JCAP0524023,JCAP0524032,2401.17689,2406.00579,2407.07416,2408.03241}.

On the other hand, the existence of singularities and their inconsistency with quantum physics suggest that general relativity (GR) might not be the ultimate theory of spacetime \cite{PRL14-57}.
This has led to a growing focus on modifying gravitational theories to incorporate quantum properties, which has become a mainstream approach in the search for the ultimate theory.
As part of this effort, Loop quantum gravity modifies spacetime near singularities, providing non-singular black hole solutions \cite{ATMP7-233}; String theory explains black hole entropy through microstates \cite{PLB379-99}; and Effective field theory introduces higher-derivative terms to the gravitational action, leading to modified metrics \cite{PRD50-3874}; etc \cite{PRL87-141601,JHEP0421220,PRL130-101501,EPJC83-83,2406.13461,
PRD109-026004,JHEP0124012,JHEP0424052,JHEP0724252}.
Very recently, a long-standing issue regarding general covariance in spherically symmetric gravity, which arises when canonical quantum gravity leads to a semiclassical model of black holes, was investigated in Ref. \cite{2407.10168}.
This work presents two distinct black hole models, distinguished by the selection of a quantum parameter.
Subsequently, using classical methods, the quasi-normal modes and the shadow radius of these two black hole models were studied in Ref. \cite{2408.02578}.

In this paper, we will first investigate the topological properties of the light rings of the two black hole solutions proposed in Ref. \cite{2407.10168} and demonstrate that these light rings are both standard and unstable according to the classification of light rings.
After that, we will discuss the shadows of these two black hole models and find that, with the increase of the quantum parameter, the shadow size of the first type of static black hole decreases, while the shadow size of the second type of static black hole remains unchanged. Additionally, the size of the lensing ring for both types of black holes decreases. The organization of this paper is outlined as follows. In Sec. \ref{II}, using the topological method \cite{PRL119-251102,PRL124-181101,PRD102-064039}, we derive the light rings of the two types of static black holes in effective quantum gravity (EQG) \cite{2407.10168}. In Sec. \ref{III}, using the geodesic equation, we derive the orbital equations for photons in the spacetime and estimate the angular radius of the supermassive black holes Sgr A* and M87* within this theoretical framework. Finally, the paper is ended up with our summaries in Sec. \ref{IV}.

%%%%%%%%%%%%%%%%%%%%%%%%%%%%%%%%%
\section{Topological properties of light rings}\label{II}
%%%%%%%%%%%%%%%%%%%%%%%%%%%%%%%%%
In this section, using the topological method \cite{PRL119-251102,PRL124-181101,
PRD102-064039}, we investigate the properties of light rings of the two types of static black holes in EQG \cite{2407.10168}, whose metrics are
\begin{equation}
\begin{aligned}
\label{metric1}
d\bar{s}^2 &= -A_1dt^2 +\frac{dr^2}{B_1} +r^2d\theta^2+r^2\sin^2\theta d\varphi^2 , \\
A_1 &= B_1 = 1 -\frac{2M}{r} +\frac{\zeta^2}{r^2}\left(1 -\frac{2M}{r} \right)^2 ,
\end{aligned}
\end{equation}
and
\begin{equation}
\begin{aligned}
\label{metric2}
d\hat{s}^2 &= -A_2dt^2 +\frac{dr^2}{B_2} +r^2d\theta^2+r^2\sin^2\theta d\varphi^2 , \\
A_2 &= 1 -\frac{2M}{r} \, , \quad B_2 = A_2\left(1 +A_2\frac{\zeta^2}{r^2}\right)  ,
\end{aligned}
\end{equation}
respectively. Here $M$ is the ADM mass, and $\zeta$ is the quantum parameter.

In order to study the topological property of the light rings, one can first introduce a potential function as \cite{PRL124-181101,PRD102-064039}
\be
H(r,\theta) = \sqrt{-\frac{g_{tt}}{g_{\varphi\varphi}}} = \frac{\sqrt{A}}{r\sin\theta} \, ,
\ee
where the function $H(r,\theta)$ is regular for $r > 2M$. Clearly, the radius of the light rings is determined by the root of $\pp_rH = 0$. According to Ref. \cite{PRD102-064039}, the key vector $\phi = (\phi^r, \phi^\theta)$ is defined as
\bea\label{vector}
\phi = \left(\frac{\pp_r H}{\sqrt{g_{rr}}} \, , ~ \frac{\pp_\theta H}{\sqrt{g_{\theta\theta}}}\right) \, .
\eea
The above vector $\phi$ can also be rewrite as
\be\label{v2}
\phi = ||\phi||e^{i\hat{\Theta}} \, ,
\ee
where $||\phi|| = \sqrt{\phi^a\phi^a}$. It is worth to note that the zero point of vector $\phi$ exactly coincides with the location of the light rings. This suggests that $\phi$ in Eq. (\ref{v2}) is not well defined for the light rings, thus one can consider the vector as $\phi = \phi^r +i\phi^\theta$. The normalized vectors are given by
\be
n^a = \frac{\phi^a}{||\phi||} \, , \quad a = 1,2\, ,
\ee
where $\phi^1 = \phi^r$, $\phi^2 = \phi^\theta$, respectively. What is more, a topological current can be established through the utilization of Duan's theory \cite{NPB514-705,
PRD61-045004} on $\phi$-mapping topological currents as follows:
\be\label{jmu}
j^{\mu}=\frac{1}{2\pi}\epsilon^{\mu\nu\rho}\epsilon_{ab}\pp_{\nu}n^{a}\pp_{\rho}n^{b}\, . \qquad
\mu,\nu,\rho=0,1,2,
\ee
where $\pp_{\nu}= \pp/\pp x^{\nu}$ and $x^{\nu}=(t,r,\theta)$. It is simple to verify that this topological current obeys
\be
\pp_{\mu}j^{\mu} = 0 \, .
\ee
This argument obviously indicates that $j^\mu$ is nonzero only at the zero points of $\phi^a(x_i)$, namely, $\phi^a(x_i) = 0$. Therefore, the topological charge at the given parameter region $\Sigma$ can be determined by using the following formula:
\be
Q = \int_{\Sigma}j^{0}d^2x = \sum_{i=1}^{N}\beta_{i}\eta_{i} = \sum_{i=1}^{N}w_{i}\, .
\ee
Here, the positive Hopf index $\beta_i$ represents the number of loops made by $\phi^a$ in the vector $\phi$-space as $x^{\mu}$ moves around the zero point $z_i$, the Brouwer degree $\eta_{i}= \mathrm{sign}(J^{0}({\phi}/{x})_{z_i})=\pm 1$. For a closed, smooth loop $C_i$
that encloses the $i$th zero point of $\phi$ while excluding other zero points, the winding number of the vector is then given by
\be
w_i = \frac{1}{2\pi}\oint_{C_i}d\Omega \, ,
\ee
where $\Omega = \arctan(\phi^\theta/\phi^r)$.

In the following, we will explore the topological property of light rings of the two types of static black holes in EQG, respectively.

%%%%%%%%%%%%%%%%%%%%%%%%%%%%%%%%%%%%%%%%%%%%%%%%%%
\begin{figure}[t]
\centering
\includegraphics[width=0.4\textwidth]{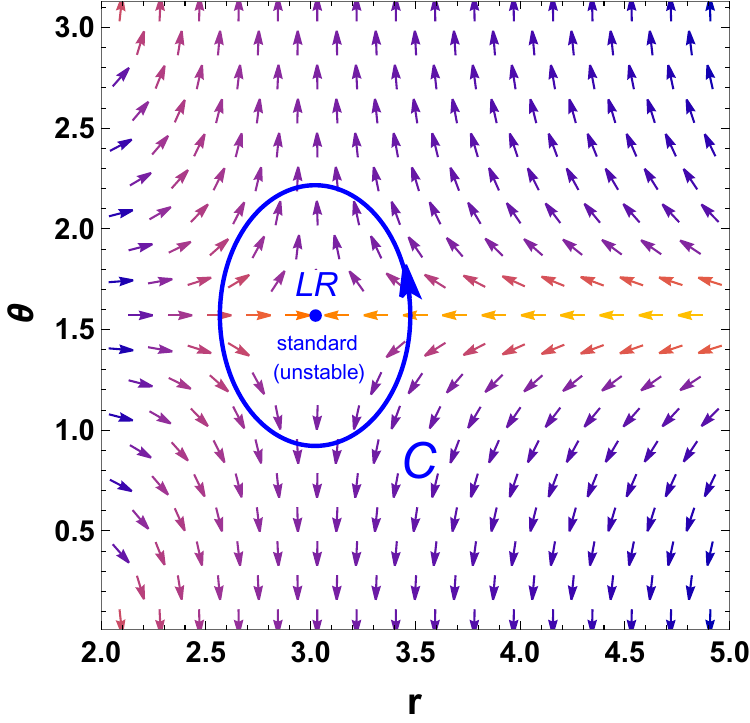}
\caption{The arrows represent the unit vector field $n$ on a portion of the $r-\Theta$ plane for the first type of static black holes in EQG with $M = 1$ and $\zeta = 1$. The light ring (LR) marked with blue dot is at $(r, \theta) = (3,\pi/2)$. The blue contour $C$ is closed loop enclosing the light ring. Obviously, the topological charge of the light ring is $Q = -1$.
\label{fig1}}
\end{figure}
%%%%%%%%%%%%%%%%%%%%%%%%%%%%%%%%%%%%%%%%%%%%%%%%%%%

For the first type of static black holes in EQG described by Eq. (\ref{metric1}), one can plot the the unit vector field $n$ on a portion of the $r-\theta$ plane in Fig. \ref{fig1} with $M = 1$ and $\zeta = 1$.
In Fig. \ref{fig1}, it is easy to observe that there is a light ring located at $(r, \theta) = (3,\pi/2)$, at the position
$r = 3M$, which is consistent with the results given in Ref. \cite{2408.02578}, obtained using the classical method \cite{PRD92-104031}.
Furthermore, The winding number $w$ for the blue contours $C$ can be characterized as $w = -1$, which is same as that for the four-dimensional Schwarzschild black hole \cite{PRL124-181101}.
Therefore, the topological charge is $Q = -1$ for the light ring of the first type of static black holes in EQG.
Based on the classification of the light rings, it is standard \cite{PRL124-181101} and unstable \cite{PRD102-064039}.

For the second type of static black holes in EQG described by Eq. (\ref{metric2}), the conclusions obtained are consistent with those of the above first type of static black holes (\ref{metric1}) in EQG, so the details are not repeated here. In other words, there is a standard and unstable light ring located at $r = 3M$ for the second type of static black holes in EQG, which is also consistent with the corresponding results given in Ref. \cite{2408.02578}.

\section{SHADOWS OF BLACK HOLES} \label{III}

\subsection{Photon orbits}\label{Sec.31}
In this section, we briefly provide an overview of photon trajectories in EQG spacetimes.
When an observer views a background source through this gravitational field, the black hole's shadow is formed by photons that are deflected and do not reach the observer.
This shadow outlines the geodesics of photons within the black hole's spacetime.
The geodesics for photon motion are defined by the Hamilton-Jacobi equation:
\begin{align}\label{EqHJ}
\frac{\pp \mathcal{S}}{\pp \tau}=-\frac{1}{2}g^{ab}\frac{\pp \mathcal{S}}{\pp x^a}\frac{\pp \mathcal{S}}{\pp x^b}.
\end{align}
The affine parameter of the null geodesic is denoted by $ \tau $, while \( \mathcal{S} \) represents the Jacobi action of the photon.
This action can be separated as:
\begin{align}\label{actionS}
\mathcal{S}=\frac{1}{2}m^2\lambda-\mathcal{E}t+L_z\varphi+\mathcal{S}_r(r)+\mathcal{S}_\theta(\theta),
\end{align}
where $ m $ is the mass of the particle (with $ m=0 $ for a photon), $ \mathcal{E} $ is the energy, and $ L_z $ is the angular momentum of the photon along the rotation axis.
The terms $ \mathcal{S}_r(r) $ and $ \mathcal{S}_\theta(\theta) $ depend only on $ r $ and $ \theta $, respectively.

If a spacetime possesses a certain symmetry, a corresponding Killing vector field can be found.
A Killing vector field $ \xi^a $ satisfies the Killing equation:
\begin{align}
\nabla_{(a}\xi_{b)} = 0.
\end{align}
This implies that the spacetime geometry remains unchanged in the direction of the Killing vector field \cite{Waldbook,Liu2023}.
For instance, if the spacetime exhibits time translation symmetry, a Killing vector field in the time direction $ \xi^a=(\pp/\pp t)^a $ exists, representing energy conservation.
Likewise, if the spacetime has axial symmetry, there is a Killing vector field $ \psi^a=(\pp/\pp \varphi)^a $, representing angular momentum conservation.
Using the properties of Killing vector fields, the conserved quantities $ \mathcal{E} $ (energy) and $ L_z $ (angular momentum) are defined as:
\begin{equation}\label{gabE}
\begin{aligned}
\mathcal{E}=&-g_{ab}\xi^a\dot{x}^b=-g_{tt}\dot{t}, \\
L_z=&g_{ab}\psi^a\dot{x}^b=g_{\varphi\varphi}\dot{\varphi}.
\end{aligned}
\end{equation}

Combining these definitions with the Hamilton-Jacobi equation \meq{EqHJ} gives the two equations of motion for photons:
\begin{align}\label{taut}
\dot{t}=\frac{\mathcal{E}}{A}, &&
\dot{\varphi}=-\frac{L_z}{r^2 \sin^2\theta}.
\end{align}
Substituting these into the null geodesic equation $ g_{ab}\dot{x}^a\dot{x}^b=0 $, we obtain:
\begin{equation}
\begin{aligned}\label{nge}
B\dot{r}^2+r^2\dot{\theta}^2&+\frac{L_z^2\csc^2\theta}{r^2}-\frac{\mathcal{E}^2}{A}=0.
\end{aligned}
\end{equation}
Employing the Carter constant $ \mathcal{K} $ \cite{Carter:1968rr} to separate variables in the above Eq. (\ref{nge}), two additional equations of motion for photons can be easily computed as:
\begin{align}\label{taur}
r^2\dot{r}=\sqrt{\mathcal{R}(r)},&&
r^2\dot{\theta}=\sqrt{\Theta(\theta)},
\end{align}
with the functions \( \mathcal{R}(r) \) and \( \Theta(\theta) \) defined as:
\begin{align}
\mathcal{R}(r)=\frac{r^4\mathcal{E}^2}{AB}-\frac{(\mathcal{K}+L_z^2)r^2}{B},&&
\Theta(\theta)=\mathcal{K}-L_z^2\cot^2\theta.
\end{align}
These equations \meq{taut} and \meq{taur} describe the propagation of light around a black hole in EQG.

To study the trajectories of photons, it is convenient to write the radial geodesics in terms of the effective potential $ V_\text{eff}(r) $ as
\begin{align}\label{Veff}
\left(\frac{\sqrt{AB}}{\mathcal{E}}\frac{dr}{d\tau}\right)^2+V_\text{eff}(r)=0,
\end{align}
with
\begin{equation}
\begin{aligned}
V_\text{eff}(r)&=-r^{-4}\mathcal{R}(r)AB/\mathcal{E}^2=-1+\frac{A(\eta+\xi^2)}{r^2},
\end{aligned}
\end{equation}
The impact parameters $\xi=L_z/\mathcal{E}$ and $\eta=\mathcal{K}/\mathcal{E}^2$ can be defined \cite{Chandrasekhar,JCAP0614043,PRD97-064021}.
The black hole's shadow silhouette is derived from the specific orbit in the radial equation, defined by $r=r_p$, which satisfies the conditions \cite{Meng:2023uws,PRD108-L041501}
\begin{align}\label{YGDTJ}
V_\text{eff}(r)\big|_{r=r_p}=0,&&\frac{d}{dr}V_\text{eff}(r)\big|_{r=r_p}=0.
\end{align}
We observe that the equations governing the shadow of the black hole depend only on the metric function $A(r)$ and are independent of $B(r)$.
This means that the quantum corrections introduced by the second type of static black holes in EQG are not reflected in the black hole shadow, making it identical to the Schwarzschild case, where the light ring radius is \( r_p = 3M \) and the shadow radius is \( R_s = 3\sqrt{3}M \). This result is consistent with the findings in the previous section and in Ref. \cite{2408.02578}.
Therefore, we now focus on the black hole spacetime corresponding to the first type of static black hole in EQG.
Subsequently, through Eqs. \meq{YGDTJ}, we can obtain the parameter equation and its first derivative equation, as shown below:
\begin{align}
&\eta+\xi^2-\frac{r_p^6}{(r_p-2M)\left[r_p^3+\zeta^2(r_p-2M)\right]}=0,\\ \label{PSs}
&\frac{2r_p^5(r_p-3M)\left[r_p^3+2\zeta^2(r_p-2M)\right]}{(r_p-2M)^2\left[r_p^3+\zeta^2(r_p-2M)\right]^2}=0.
\end{align}
Here, the real root of equation \meq{PSs} represent the light ring radius, which is $ r_p=3M $, remaining consistent with the Schwarzschild case.
And then, we can obtain the parameter equation at the light ring radius,
\begin{align}
\eta+\xi^2=729M^4/(27M^2+\zeta^2).
\end{align}

\subsection{Apparent shape}\label{Sec41}

When photons from a light source pass near a black hole, they are deflected by its strong gravitational field, a process known as gravitational lensing.
Some of these photons reach a distant observer, while others are captured by the black hole.
The captured photons create the black hole's shadow, a dark region in the observer's sky where the gravitational pull is so strong that light cannot escape.
Since the effective potential for photon motion in the second type of static black hole spacetime mirrors that of Schwarzschild spacetime, this subsection will focus solely on the first type of static black holes.
We opt to utilize the following normalized and orthogonal tetrad \cite{Zhang:2020xub}:
\begin{equation}\label{zjbj}
\begin{aligned}
e_{(t)}=&\left.\frac{1}{\sqrt{-g_{tt}}}\pp_t\right|_{(r_0,\theta_0)},\\
e_{(r)}=&\left.-\frac{1}{\sqrt{g_{rr}}}\pp_r\right|_{(r_0,\theta_0)},\\
e_{(\theta)}=&\left.\frac{1}{\sqrt{g_{\theta\theta}}}\pp_\theta\right|_{(r_0,\theta_0)},\\
e_{(\varphi)}=&-\left.\frac{1}{\sqrt{g_{\varphi\varphi}}}\pp_\varphi\right|_{(r_0,\theta_0)},
\end{aligned}
\end{equation}
where the observer is located at $ (r_0,\theta_0) $ in the coordinates $ \{t,r,\theta,\varphi\} $ and $ g_{ab} $  represents the metric component of the background spacetime.
It's important to understand that the choice of tetrad isn't unique and can be tailored to specific needs.
Although light typically travels from the source to the observer in practical situations, for calculations, we can assume it originates from the observer due to the reversibility of optical paths.
The four-momentum components, using an orthogonal tetrad, are expressed as \( p^{(t)} = -p_\mu e^{\mu}_{(t)} \) and \( p^{(i)} = p_\mu e^{\mu}_{(i)} \), where \( (i) = (r, \theta, \varphi) \).
This represents the four-momentum as measured by a locally static observer.
For a massless photon, the three-vector linear momentum \(\vec{P}\) relates to \(p^{(i)}\) and satisfies \(|\vec{P}|=p^{(t)}\) in the observer's frame. The observation angles \((\alpha,\beta)\) are defined as:
\begin{equation}
\begin{aligned}
p^{(r)}&=|\vec{P}|\cos\alpha \cos\beta, \\
p^{(\theta)}&=|\vec{P}|\sin\alpha, \\
p^{(\varphi)}&=|\vec{P}|\cos\alpha\sin\beta
\end{aligned}
\end{equation}
These angles specify the direction of light rays in the observer's local sky. The coordinates $ (x,y) $ in the local sky are related to \((\alpha,\beta)\) by:
\begin{equation}
\begin{aligned}
x&=-r_0\tan\beta=-r_0\frac{p^{(\varphi)}}{p^{(r)}}, \\
y&=r_0\frac{\tan\alpha}{\cos\beta}=r_0\frac{p^{(\theta)}}{p^{(r)}}
\end{aligned}
\end{equation}
The black hole's shadow in the observer's sky, composed of pixels corresponding to light rays that fall into the black hole, is defined by the boundary of unstable spherical photon orbits.

In a static black hole spacetime in EQG, the observed position of the photon's image in the sky is determined by:
\begin{equation}
\begin{aligned}
x &= -r_0 \frac{p^{(\varphi)}}{p^{(r)}} = -\frac{r_0\xi \csc\theta_0 }{\sqrt{r_0^2/A(r_0)-\eta-\xi^2}}, \\
y &= r_0 \frac{p^{(\theta)}}{p^{(r)}} =  \frac{r_0\sqrt{\eta -\xi^2\cot^2\theta_0} }{\sqrt{r_0^2/A(r_0)-\eta-\xi^2}},
\end{aligned}
\end{equation}
assuming the observer is located at a distance $r = r_0$ and $\theta = \theta_0$.
Considering the static black hole in asymptotically flat spacetime, the position of a real observer can be taken at infinity, allowing us to take the limit $r_0 \to \infty$, which yields:
\begin{align}\label{XXX}
x= -\xi \csc \theta_0, &&  y= \sqrt{\eta - \xi^2 \cot^2 \theta_0}.
\end{align}
We can easily observe that the celestial coordinates satisfy:
\begin{align}\label{X2Y2}
x^2+y^2=\eta+\xi^2=729M^4/(27M^2+\zeta^2).
\end{align}
To visualize the impact of quantum parameter on the black hole shadows, we have plotted the shadow comparison diagrams for different parameters.
Fig. \meq{fig2} shows that as the quantum parameter increases, the black hole shadow contours become smaller.
\begin{figure}[t]
\centering
\includegraphics[width=0.35\textwidth]{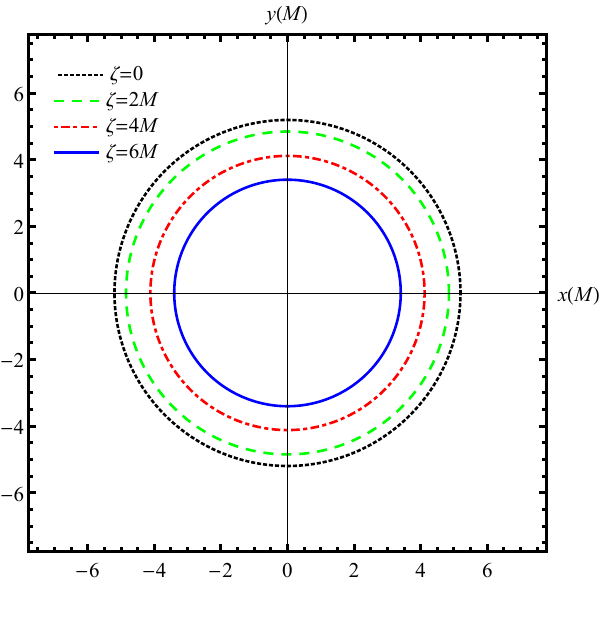}
\caption{Shadow contours under different quantum parameters for the first type of static black holes in EQG.}
\label{fig2}
\end{figure}

For a approximatively estimation, utilizing the metric \meq{metric2}, we calculate the angular radius of a black hole shadow, defined as $ \theta_{\text{BH}} = R_s \frac{\mathcal{M}}{D_O} $, with $D_O$ representing the distance from the observer to the black hole.
Here, according to equation \meq{X2Y2}, the black hole shadow radius is
\begin{align}
R_s=\sqrt{x^2+y^2}= \frac{27M^2}{\sqrt{\zeta^2+27M^2}}.
\end{align}
Specifically, for a black hole with mass $ \mathcal{M} $ located at a distance $ D_O $ from the observer, the angular radius $ \theta_\text{BH} $ can be expressed as \cite{Amarilla:2011fx}
\begin{align}
\theta_\text{BH} = 9.87098 \times 10^{-6} \frac{27\mathcal{M}}{\sqrt{27+\tilde{\zeta}^2}M_\odot}  \left(\frac{1 \text{kpc}}{D_O}\right) \mu\text{as},
\end{align}
where $ \tilde{\zeta}=\zeta/M $ is the dimensionless quantum parameter.
\begin{figure}[t]
\centering
\includegraphics[width=0.4\textwidth]{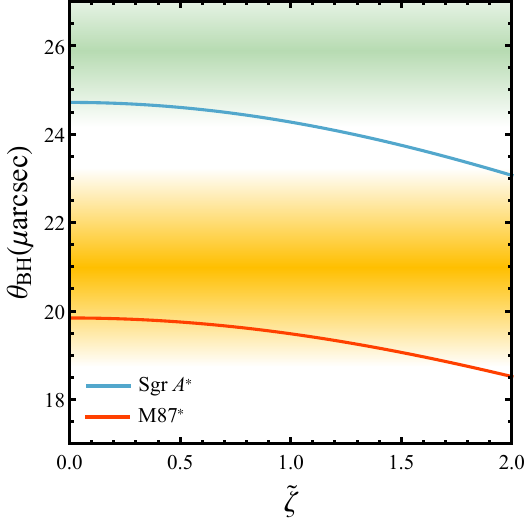}
\caption{The estimation of the angular radius of the supermassive black holes Sgr A* and M87* using the metric of the first type of static black hole in EQG. }
\label{fig3}
\end{figure}
Fig. \ref{fig3} presents the calculated angular radius of the black holes at the Galactic center (Sgr A*) and in the galaxy M87 (M87*) using a static black hole metric in EQG.
The latest observations give Sgr A* a mass of \(\mathcal{M} = 4.0 \times 10^6 M_\odot\) and a distance of \(D_O = 8.3\) kpc from the observer \cite{APJL930-L12}.
For M87*, the mass is \(\mathcal{M} = 6.5 \times 10^9 M_\odot\) with an observer distance of \(D_O = 16.8\) Mpc \cite{APJL875-L6}.
It is important to note that we used different gradient colors in the background of Fig. \ref{fig3} to broadly highlight the angular radius data for the Sgr A* and M87* black holes, which are within the general range of recent observations.
\footnote{The angular diameters of the black holes Sgr A* and M87* have been measured as  $51.8 \pm 2.3 \mu\text{as}$ \cite{APJL930-L12} and $42 \pm 3.0 \mu\text{as}$ \cite{APJL875-L1}, respectively.}
The gradient green and gradient yellow correspond to the Sgr A* and M87* black holes, respectively.

Although the supermassive black holes Sgr A* and M87* in real astronomical environments are both rotating, the spin parameter significantly affects the distortion parameter of the shadow while having little effect on the angular radius of the shadow \cite{2406.00579}.
This implies that it is reasonable to constrain the quantum parameters using static EQG black holes before obtaining rotating EQG black holes.
In Table \ref{tab1}, we list the percentage deviation in the angular radius of the shadow between Kerr black holes and Schwarzschild black holes as the dimensionless spin parameter $ a/M $ increases, where the percentage deviation is defined by
\begin{align}
\delta=\frac{R_s^\text{Kerr}-R_s^\text{Sch.}}{R_s^\text{Sch.}}\times 100\%,
\end{align}
and the definition of $R_s^\text{Kerr}$ is provided in Fig. 3 of Ref. \cite{Hioki:2009na}.
\begin{table}[t]
\renewcommand{\arraystretch}{1.25}
\centering
\setlength\tabcolsep{1.3mm}{
\begin{tabular}{cccccccccc}
\hline\hline
\multirow{2}{*}{$ a/M $} & \multirow{2}{*}{$ 0.2 $}& \multirow{2}{*}{$ 0.4 $}& \multirow{2}{*}{$ 0.6 $}& \multirow{2}{*}{$ 0.8 $} & \multirow{2}{*}{$ 0.99 $}
\\ \\
\hline\\
  $ \delta $    &0.0003\%~~   & 0.0028\%~~    &0.0122\%~~      &0.0341\%~~    &0.0716\%~~   \\ \\
\hline\hline
\end{tabular}}
\caption{The deviation in the angular radius of the shadow between Kerr black holes with different spin parameters and Schwarzschild black holes is analyzed.}
\label{tab1}
\end{table}

\subsection{Lensing rings}\label{Sec43}

In the previous section, we analytically analyzed the light rings and shadows of two types of static black holes in EQG, finding that their light rings are identical to the Schwarzschild case, and the shadow of the second type of black hole is also identical to the Schwarzschild case.
A very natural question arises: how can we distinguish EQG from GR?
This question becomes particularly challenging for the second type of static black hole, where quantum corrections do not manifest in the shadow.
To address this, we will adopt the numerical backward ray-tracing method \cite{PRD103-044057} to investigate the black hole shadow.
This method can visually represent the proportion of the shadow in the lensing ring and has the potential to become a criterion for distinguishing EQG from GR.

\begin{figure}[t]
\centering
\includegraphics[width=0.8\linewidth]{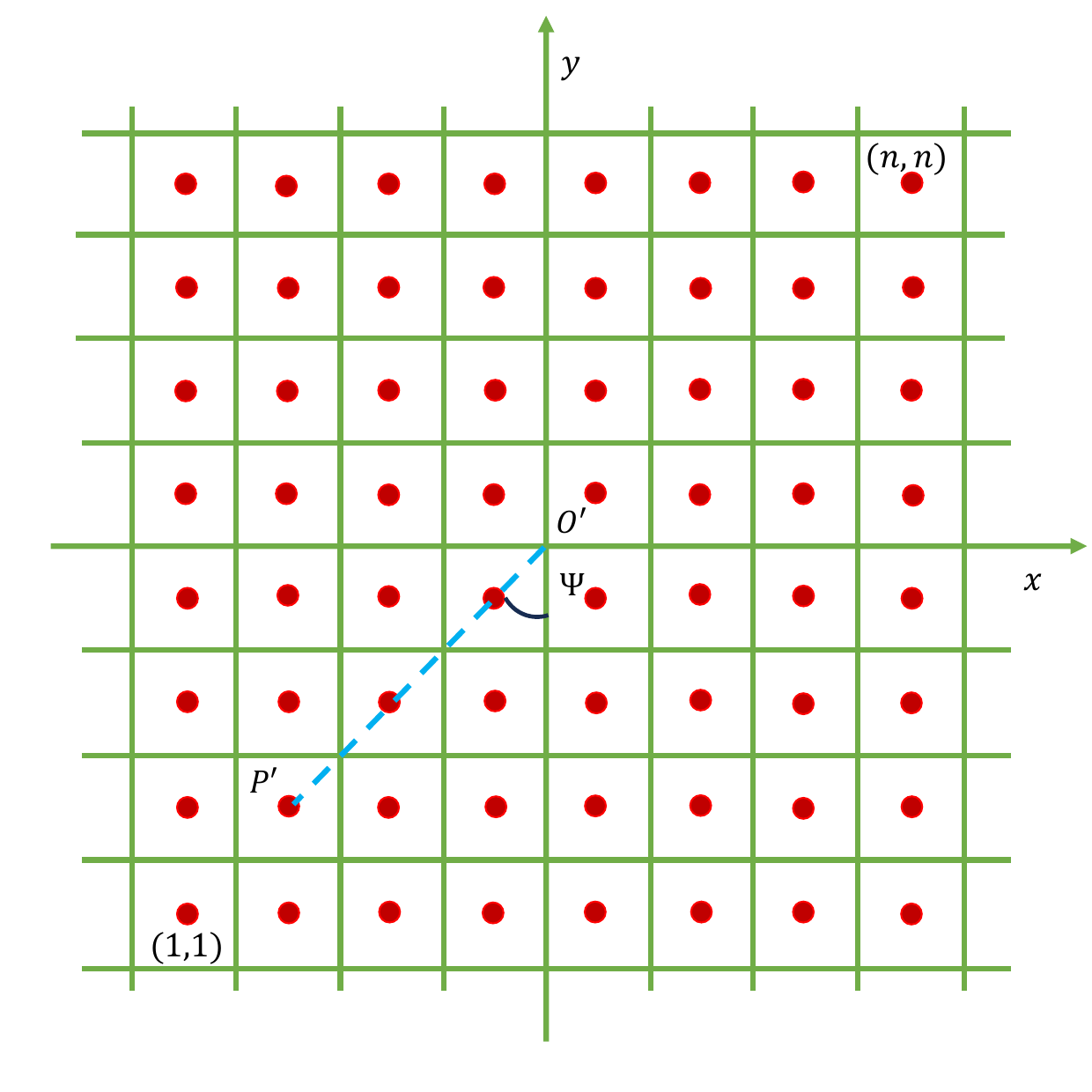}
\caption{
Illustration of the pixels.
A standard Cartesian coordinate with the origin $ O' $ is set up. This diagram has been shown in Fig. 9 of our previous work \cite{2406.00579}.}
\label{fig4}
\end{figure}

In the selected tetrad \meq{zjbj}, $e_0$ denotes the observer's four-velocity, $e_3$ points towards the black hole's center, and the combination $e_0 \pm e_3$ aligns with the principal null directions of the metric.
For each light ray $ \lambda(s) $ with coordinate representation $ t(s), r(s),\vartheta(s),\varphi(s) $, the general form of tangent vector is
\begin{align}\label{lambda1}
\dot{\lambda}=\dot{t}\pp_t+\dot{r}\pp_r+\dot{\vartheta}\pp_\vartheta+\dot{\varphi}\pp_\varphi.
\end{align}
In the observer's frame, the tangent vector of the null geodesic can be expressed as
\begin{equation*}
\dot{\lambda}=|\overrightarrow{OP}|(- \chi e_0+\sin\theta\cos\psi e_1+\sin\theta\sin\psi e_2+\cos\theta e_3),
\end{equation*}
where $\chi$ is a scalar factor, and $ |\overrightarrow{OP}| $ denotes the tangent vector of the null geodesic at the point $ O $ in the three-dimensional space.
The stereographic projection from the celestial sphere onto a plane is then re-obtained
\begin{equation}\label{B4}
\begin{aligned}
x_{P'}=&-2|\overrightarrow{OP}|\tan\frac{\theta}{2}\sin\psi,\\
y_{P'}=&-2|\overrightarrow{OP}|\tan\frac{\theta}{2}\cos\psi,
\end{aligned}
\end{equation}
where $ P' $ is the projection of point $ P $ onto the Cartesian plane.
A standard Cartesian coordinate system with origin $ O' $ is set up as shown in Fig. \ref{fig4}.
\begin{figure}[t]
\centering
\includegraphics[width=0.8\linewidth]{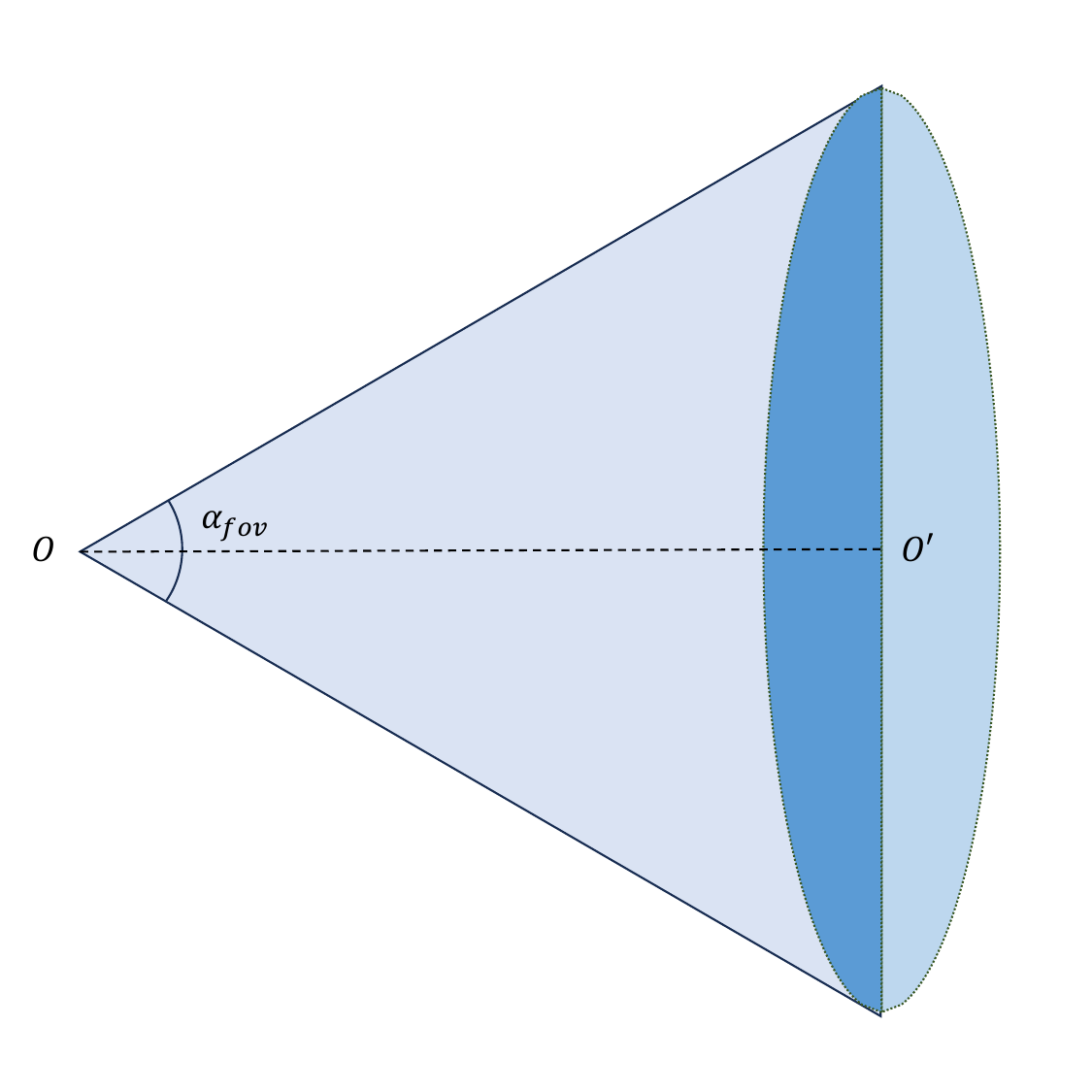}
\caption{The field of view.}
\label{fig5}
\end{figure}
Next, we discuss the field of view, defined by the angles in the $yO'z$ and $xO'z$ planes, assumed equal for simplicity.
Figure \ref{fig5} illustrates an example.
The length $L$ of the square screen is given by:
\begin{align}
L=2|\overrightarrow{OP}|\tan\frac{\alpha_\text{~fov}}{2}.
\end{align}
We consider the screen with $n \times n$ pixels, where the length occupied by each pixel is given by
\begin{align}
\ell=\frac{2|\overrightarrow{OP}|}{n}\tan\frac{\alpha_\text{~fov}}{2}.
\end{align}
The pixels are indexed by $(i, j)$, with $(1,1)$ representing the bottom left corner and $(n,n)$ the top right corner, where both $ i $ and $ j $ range from $ 1 $ to $ n $.
Then, the Cartesian coordinates for the center point of a pixel can be given by
\begin{align}\label{B7}
x_{P'}=\ell \left(i-\frac{n+1}{2}\right), && y_{P'}=\ell\left(j-\frac{n+1}{2}\right).
\end{align}
By comparing Eqs. \meq{B4} and \meq{B7}, we can derive the relationship between the pixel indices $ (i,j) $ and the angles $ (\theta,\Psi) $ as follows:
\begin{equation}
\begin{aligned}
\tan\Psi&=\frac{j-(n+1)/2}{i-(n+1)/2},\\
\tan\frac{\theta}{2}&=\tan\frac{\alpha_\text{~fov}}{2}\frac{\sqrt{[i-(n+1)/2]^2+[j-(n+1)/2]^2}}{n}.
\end{aligned}
\end{equation}

\begin{figure}[t]
\centering
\includegraphics[width=0.9\linewidth]{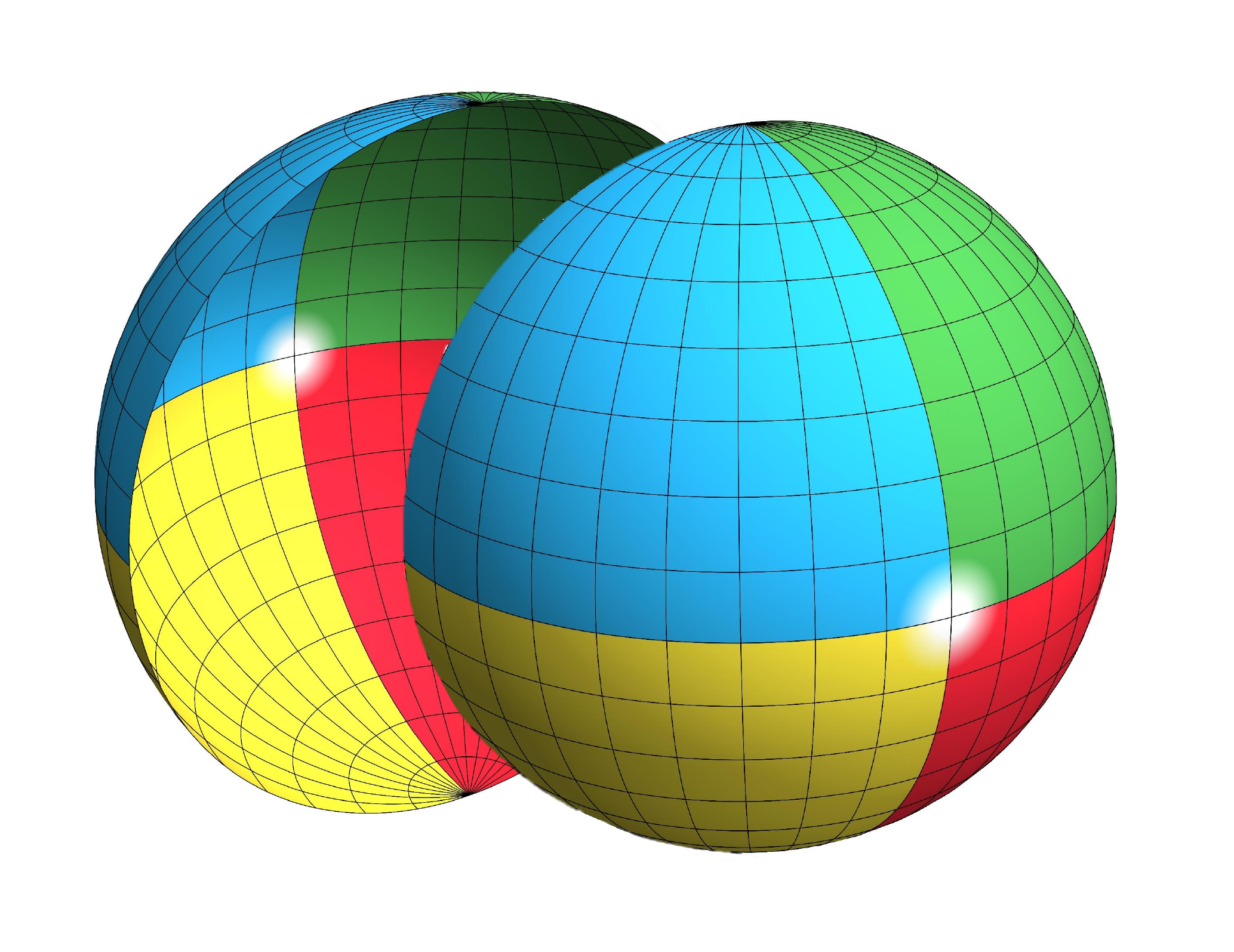}
\caption{Illustration of our spherical light source at infinity.}
\label{fig6}
\end{figure}

\begin{figure}[t]
\centering
\includegraphics[width=0.48\linewidth]{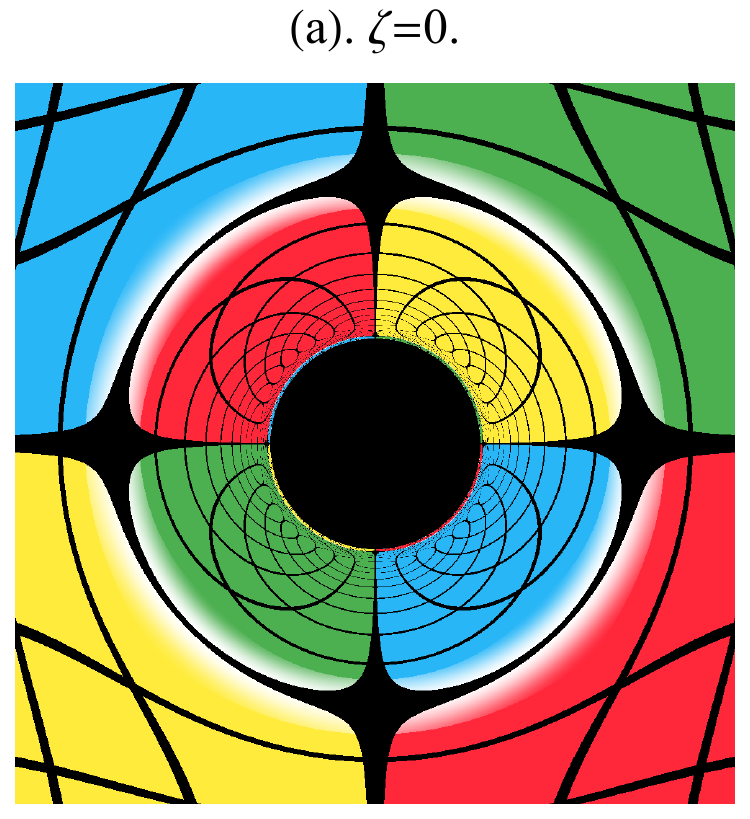}
\includegraphics[width=0.48\linewidth]{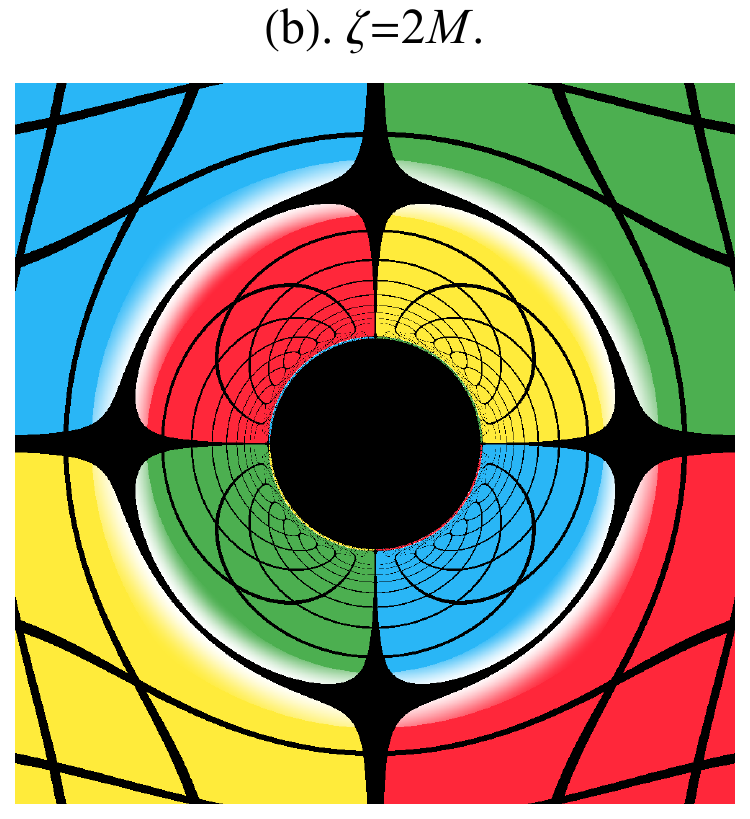}
\includegraphics[width=0.48\linewidth]{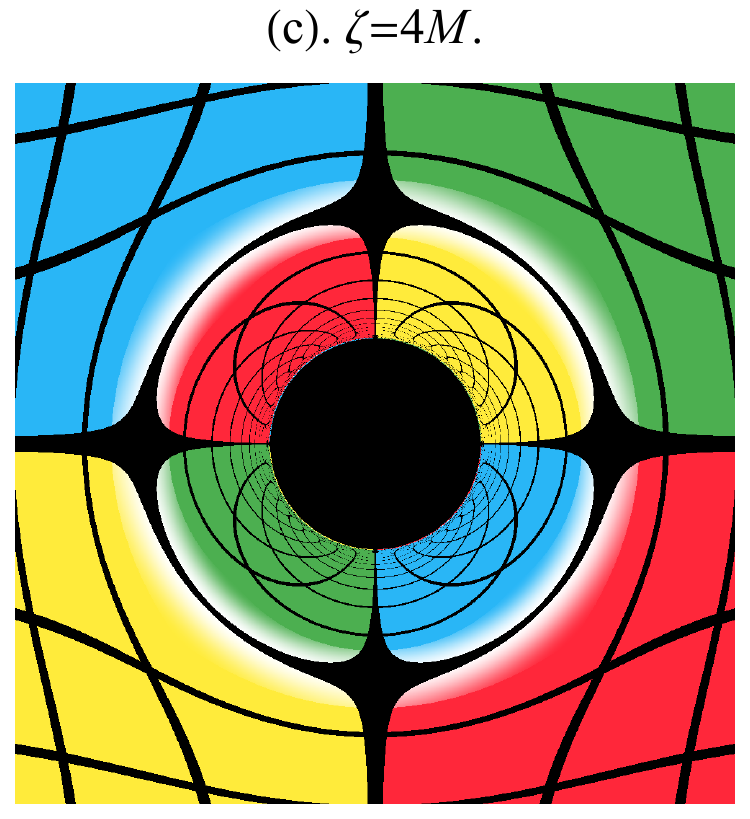}
\includegraphics[width=0.48\linewidth]{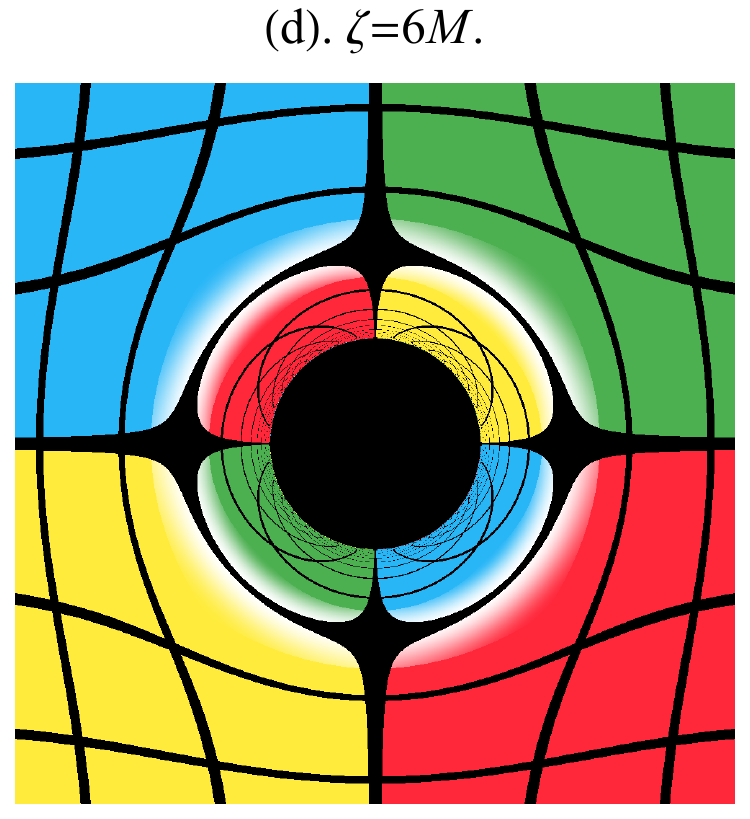}
\caption{Shadows and lensing rings of the second type of static black hole in EQG with a quantum parameter $\zeta$, as seen by an observer at $\theta_0 = \pi/2$.}
\label{fig7}
\end{figure}

\begin{figure}[t]
\centering
\includegraphics[width=0.48\linewidth]{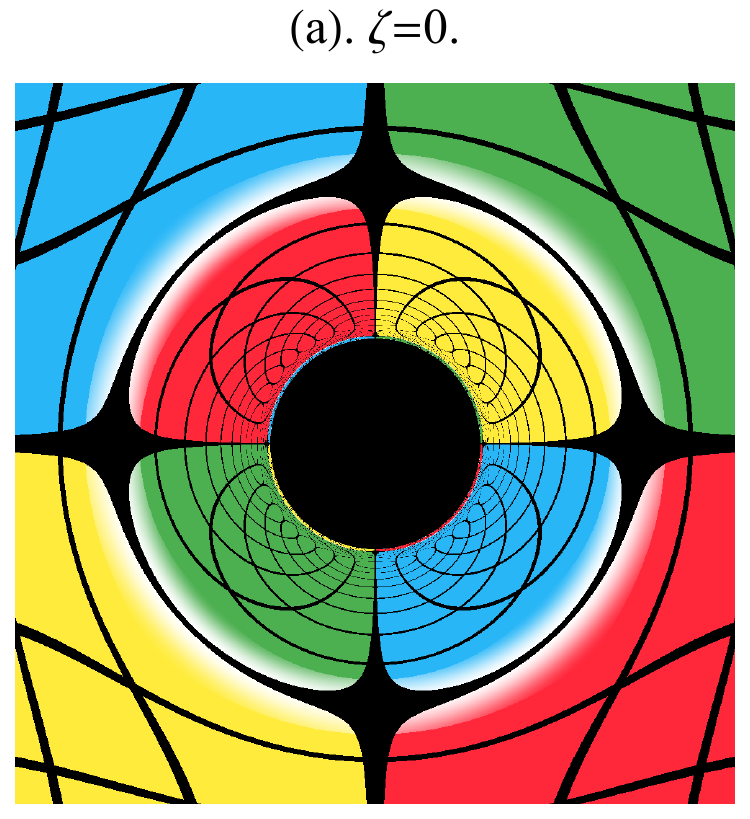}
\includegraphics[width=0.48\linewidth]{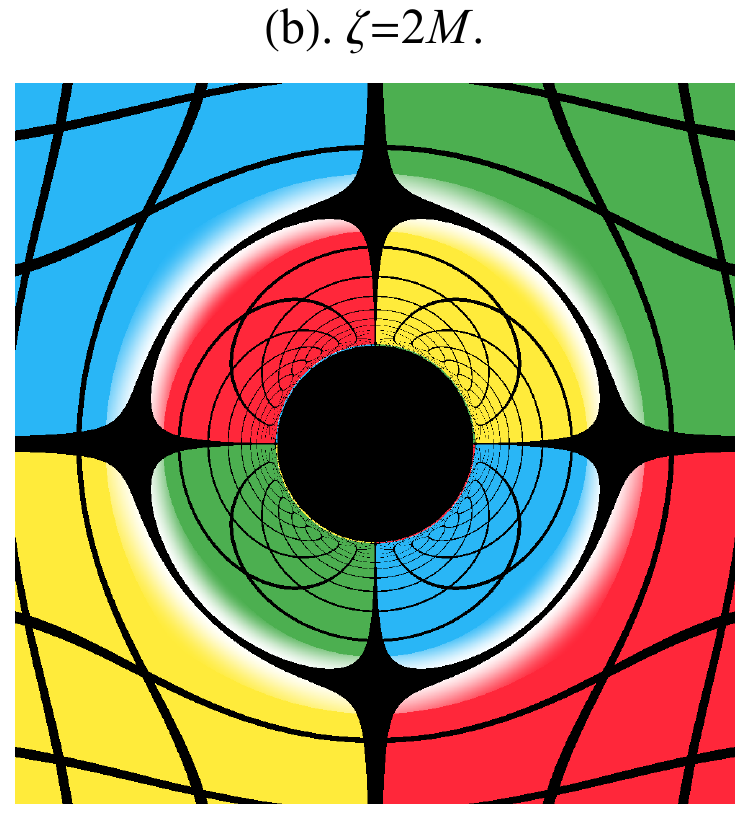}
\includegraphics[width=0.48\linewidth]{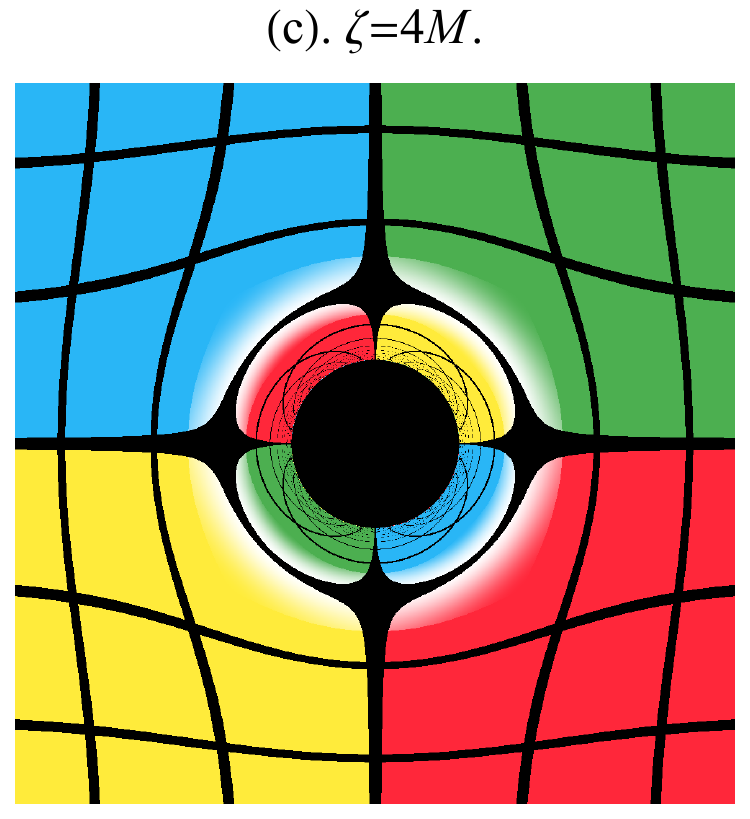}
\includegraphics[width=0.48\linewidth]{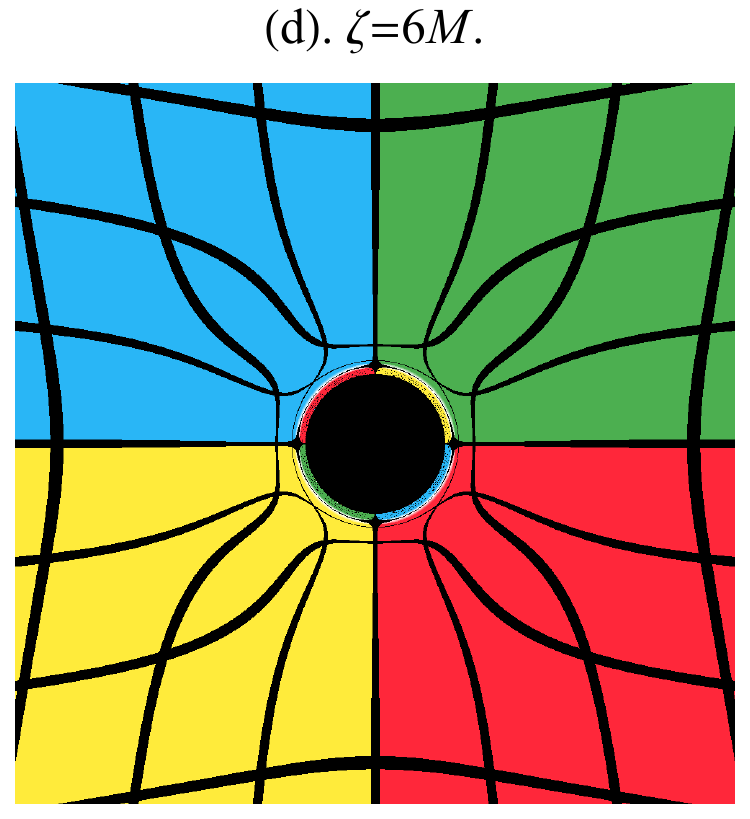}
\caption{Shadows and lensing rings of the first type of static black hole in EQG with a quantum parameter $\zeta$, as seen by an observer at $\theta_0 = \pi/2$.}
\label{fig8}
\end{figure}

Before presenting the results, it is essential to note that an extended light source was used to illuminate the system.
Figure \ref{fig6} illustrates a cross-sectional view of the "large sphere" that encompasses both the black hole and the observer, exposing its internal structure.
The sphere is divided into a grid of latitude and longitude lines, with adjacent lines spaced by an interval of $\pi/18$.
To generate the black hole image, each grid segment is assigned a color. The extended source is divided into four colors, and the color of each pixel in the image is determined by tracing the photons' paths from the corresponding points on the source.
The point at the origin is marked with white.
If the effect of the strong gravitational field around the black hole on the bending of light is considered, this point will appear as a lensing ring in the observed sky.
Dark areas indicate that photons have been absorbed by the black hole.

Now, we discuss the impact of the quantum parameter $\zeta$ on the black hole shadow and lensing rings.
It is important to note that we first examine the properties of the second type of black hole in EQG, as it cannot be distinguished from a Schwarzschild black hole based solely on its shadow.
As shown in Figs. \ref{fig7}, identical shadows are observed across different parameters, consistent with the analytical analysis in the previous subsection.
However, the white halo representing the lensing ring shrinks as the parameter increases, leading to the shadow occupying a larger proportion of the ring.
Then, Figs. \ref{fig8} shows the images of the first type of black hole in EQG.
With sufficiently small quantum parameters, the images of the two types of black holes are quite similar, as shown by comparing Figs. \ref{fig7}(a)-\ref{fig7}(b) with Figs. \ref{fig8}(a)-\ref{fig8}(b).
However, for sufficiently large quantum parameters, the first type of black hole in EQG exhibits rather peculiar properties, with the lensing ring nearly overlapping the shadow's contour.

\medskip

%\medskip

\section{Conclusions}\label{IV}

In the context of Loop quantum gravity, research on black holes seeks to merge quantum mechanics with general relativity to resolve challenges like black hole singularities and the information paradox, and to investigate the quantum nature of spacetime.
A recent investigation \cite{2407.10168} tackled a long-standing challenge related to general covariance in spherically symmetric gravity within semiclassical black hole models derived from EQG, introducing two different black hole models determined by different quantum parameter choices.

In this paper, via the topological approach, we firstly investigate the light rings of the two types of static black holes in EQG \cite{2407.10168}, and demonstrate that these light rings in both types of static black holes are standard and unstable according to the classification of light rings.
We then analyzed the positions of the light rings using the photon trajectory equation.
We also found that, although the quantum parameters do not alter the light rings of the two black hole types, they do contract the size of the shadow of the first type of static black hole in EQG.
In contrast, the second type of static black hole in EQG remains indistinguishable from a Schwarzschild black hole based on its shadow. Notably, the quantum parameters shrink the lensing rings for both black hole types in EQG, enlarging the proportion of the black hole shadow within the ring. This difference offers a potential way to distinguish whether the black hole is described by EQG or GR. We believe that the findings of this paper will contribute to the development of loop quantum gravity theory and help advance research on addressing key issues such as black hole singularities and the information paradox.

\acknowledgments

We are greatly indebted to the anonymous referee for the constructive comments to improve the presentation of this work. This work is supported by the National Natural Science Foundation of China (NSFC) under Grants No. 12205243, No. 12375053, No. 12122504, and No. 12035005; the Sichuan Science and Technology Program under Grant No. 2023NSFSC1347; the Doctoral Research Initiation Project of China West Normal University under Grant No. 21E028; the science and technology innovation Program of Hunan Province under Grant No. 2024RC1050; the innovative research group of Hunan Province under Grant No. 2024JJ1006; and the Hunan Provincial Major Sci-Tech Program under Grant No.2023ZJ1010; and Postgraduate Scientific Research Innovation Project of Hunan Province (CX20240531).

\end{document}